\documentclass[twocolumn,prb,aps,showpacs,preprintnumbers,amsmath,amssymb,superscriptaddress]{revtex4}
\usepackage{graphicx}
\usepackage{bm}
\begin{document}

\title{Shifted COCG method and its application to double orbital 
extended Hubbard model} 
\author{Susumu Yamamoto}
\affiliation{Core Research for Evolutional Science and Technology,
Japan Science and Technology Corporation (CREST-JST), Japan}
\author{Tomohiro Sogabe}
\affiliation{Department of Computational Science and Engineering,
Nagoya University, Furo-cho, Chikusa-ku,Nagoya 464-8603, Japan}
\author{Takeo Hoshi}
\affiliation{Department of Applied Mathematics and Physics,
Tottori University,4-101 Koyama-Minami, Tottori, 680-8550 Japan}
\affiliation{Core Research for Evolutional Science and Technology,
Japan Science and Technology Corporation (CREST-JST), Japan}
\author{Shao-Liang Zhang}
\affiliation{Department of Computational Science and Engineering,
Nagoya University, Furo-cho, Chikusa-ku,Nagoya 464-8603, Japan}
\author{Takeo Fujiwara}
\affiliation{Center for Research and Development of Higher Education,
The University of Tokyo, Tokyo 113-8656, Japan}
\affiliation{Core Research for Evolutional Science and Technology,
Japan Science and Technology Corporation (CREST-JST), Japan}

\begin{abstract}
We explains the shifted COCG method
which can solve a series of the linear equations
generated by numbers of scaler shifts,
without time consuming matrix-vector operations,
except at the only one reference energy.
This is a family of the CG method and sharing
the robustness and the capability of
the accuracy estimation.
Then shifted COCG is quite useful to calculate the Green's function
of the many-electron Hamiltonian which have very large dimension.
We applied it to the double orbital extended
Hubbard model with twelve electrons on the periodic
$\sqrt{8}\times\sqrt{8}$ site system, the dimension
of the Hamiltonian equals to 64,128,064, and found
the ground state is insulator.
We also explained the crucial points of the shifted
COCG algorithm for reducing the amount of required memory.
\end{abstract}
\pacs{71.15.Dx,71.27.+a,71.10.Fd,02.60.Dc}
\date{\today}
\maketitle

\newcommand{\STACK}[2]{\genfrac{}{}{0pt}{1}{#1}{#2}}

\section{Introduction}\label{sec_intro}

Strongly interacting systems attract considerable attention because of
fruitful phenomena in a new field of {\it cross-correlation} physics~\cite{Tokutra}
and their potential applicability 
to developing field of spintronics. 
Theoretical study of strongly correlated systems, 
e.g. many-electron systems and interacting spin systems, 
becomes time-consuming and more difficult  
when one  starts numerical investigation of larger systems. 

One reason of this difficulty is, of course, the large dimension of 
the Hilbert space or the Hamiltonian matrix of many-electron systems. 
The dimension of the Hilbert space grows exponentially 
with increasing number of atoms linearly in a many-electron system 
and, on the contrary, that in a one electron problem (or the 
density functional theory, DFT) the size of the Hamiltonian matrix 
is proportional to the number of atoms.
The second reason is 
the fact that the rigorousness or accuracy control becomes seriously difficult 
in a problem of the large Hamiltonian matrix. 
Because the width of the spectra is in proportion to the number of atoms in many cases, 
the energy interval between adjacent eigenenergies becomes small quite rapidly 
with increasing number of atoms.
The short interval between adjacent eigenenergies causes the difficulty in
separating of respective eigenvectors.
Then, for example, it is very important to obtain the precise ground state,
from which all the physical quantities are derived
in the (zero-temperature) many-electron theory.
Thus, one needs higher energy resolution with increasing number of atoms,
but, sometimes, we do not have fast, reliable and stable
calculation algorithm for large Hamiltonian matrices.

Our main target is the calculation of the Green's function matrix $G(\omega)$
in many-electron problems;
\begin{eqnarray} 
G_{ij}(\omega) = [(\omega + {\rm i}\eta -H)^{-1}]_{ij}  ,
\label{eq:gf}
\end{eqnarray}
where $H$ and $\omega$ are a real Hamiltonian matrix and energy parameter, respectively.
The suffices $i$ and $j$ denote arbitrary state such as
$\hat{c}_i \left|\right\rangle$ or $\hat{c}_i^\dagger \left|\right\rangle$,
where $\hat{c}$ is an annihilation operator and  $\left|\right\rangle$ is a ground state.
Here, we should use a positive finite parameter $\eta$
in a numerical calculation of a finite system, 
instead of infinitesimally small positive number.
The spectral function 
is the important physical quantity derived from the Green's function Eq.~(\ref{eq:gf}).

There are two possibilities for calculating Eq.~(\ref{eq:gf}).
One is to solve the eigenvalue problem with an eigenvalue $\omega$,
e.g. the Lanczos method.
The other is to solve following linear equation and to take inner product between
 the solution and vector $\left|i\right\rangle$,
\begin{eqnarray}
     && A \stackrel{\rm def}{=}\omega+{\rm i}\eta -H,
		\label{eq:A} \\
     && A \left|x_j\right\rangle = \left|j \right\rangle ,
		\label{eq:lin_eq}\\
     && G_{ij}(\omega)=\left\langle i\right.\left|x_j\right\rangle ,
		\label{eq:gf2}
\end{eqnarray}
with an arbitrary energy parameter $\omega$, 
e.g. the shifted COCG (conjugate-orthogonal-conjugate-gradient) method,
a family of the CG (conjugate-gradient) method.
In both cases, we first restrict the space dimension of states to be finite. 
In other words, we assume the size of the Hamiltonian matrix to be finite. 
Then we construct the Krylov subspace defined as
\begin{eqnarray}
{\cal K}_{n}(A, |j\rangle)={\rm span} \{|j\rangle,A|j\rangle,A^2|j\rangle,\ldots, A^n|j\rangle\}.
\label{eq:Krylov}
\end{eqnarray}

In the Lanczos method, orthogonalized base vectors (Lanczos vectors) are
successively generated in ${\cal K}_{n}(A,|j\rangle)$, and 
at the same time, the Hamiltonian matrix is tridiagonalized.
In a large scale calculation, 
one can only use a small Krylov subspace, 
because of  heavy load of computation 
and a corruption of the orthogonality of generated basis vectors. 
It is well known that the rounding error breaks down the
orthogonality of the generated base vectors rapidly,
when the dimension of the Krylov subspace exceeds several tens. 
The corruption of the orthogonality causes spurious eigenvalues and, 
more seriously, incorrect eigenvectors. 
Therefore, the size of the Krylov subspace 
should be limited usually to some tens or a hundred.

We developed the shifted COCG method,
where the Eq.~(\ref{eq:gf2}) is solved within the Krylov subspace,
and applied it to the one-electron tight-binding Hamiltonian
in the system with a large number of atoms.~\cite{Takayama}
A set of orthogonal base is created by the iterative
process of the shifted COCG method, like Lanczos process,
but the calculation is stable for 
large dimension of the Krylov subspace,
in contrast to the Lanczos method.
We must solve the Eq.~(\ref{eq:gf2}),
for every scalar shift $\sigma$ of $A$ corresponding to
respective energy mesh point.
The number of the $\sigma$'s is as much as O($10^2$)$\sim$O($10^4$)
generally,
however, the most time-consuming matrix-vector operations are
needed only at a single reference energy ($\sigma=0$).
Then the order of the total amount of calculation is just the same
as Lanczos method.
The reduction of the matrix-vector operation at non-zero $\sigma$
are based on the fact that a power of $(A+\sigma)$
is decomposed into a linear combination of powers of $A$.
Thus, Krylov subspace is invariant
${\cal K}_{n}(A,\left|j\right\rangle)
={\cal K}_{n}(A+\sigma,\left|j\right\rangle)$ against $\sigma$.

In the application of this method to the many-electron theory,
because the dimension of the vectors is huge,
we must take care for the total amount of
base vector storage for ${\cal K}_{n}(A,\left|j\right\rangle)$,
in order to satisfy the memory constraint in modern computers.
We explain the innermost loop index should be the iteration step $n$,
for an extremely large size of the Hamiltonian matrix.
This structure also give us following additional two merits.
One is that a part of the program code can be used
in the inverse iteration process to improve the ground state.
Another is that the calculation with the different $\eta$ can be done
without time consuming matrix-vector operations.

The structure of the paper is as follows.
In Sec.~\ref{Sec:sCOCG}, the basics of the shifted COCG method
is explained briefly.
Section~\ref{Sec:error_and_seed_switching}
is devoted to explanation of how to obtain global convergence.
Then we apply the shifted COCG method to an
extended Hubbard Hamiltonian with orbital degeneracy
and intra- and inter-site Coulomb interactions 
in Sec.~\ref{Sec:Application}, where
the size of Hamiltonian matrix is equal to 64,128,064.
We  calculate one-electron excitation spectra and evaluate the insulating gap. 
In Sec.~\ref{Sec:Discussion},  
we will conclude that 
the essential difficulties of numerical investigation
of many-electron problems, the accuracy control (or monitoring)
and the robustness are achieved by the present method,
within the moderate amount of memory space.
We explain the two points to understand
the mathematics in the back ground of the shifted COCG method
in Appendix~\ref{app:math}.
The practical design of storing the huge Hamiltonian matrix
is discussed in Appendix~\ref{app:storage}.

\section{Shifted COCG method}
\label{Sec:sCOCG}

Assuming that the Hamiltonian is represented by using $N$-dimensional real matrix $H$
and $A$ is a complex symmetric matrix $\omega_{\rm ref}+{\rm i}\eta_{\rm ref}-H$,
we should solve the linear simultaneous equation of
\begin{equation}
	A \bm{x}=\bm{b},
\label{Eq:ref}
\end{equation}
and its shifted equation
\begin{equation}
	(A+\sigma) \bm{x}^{\sigma}=\bm{b}, 
\label{Eq:shifted}
\end{equation}
where  $\sigma=(\omega+{\rm i}\eta)-(\omega_{\rm ref}+{\rm i}\eta_{\rm ref})$.
We represent quantities $q$ in the shifted system as $q^\sigma$.
The right hand side $\bm{b}$ represents $\left|j\right\rangle$
in Eq.~(\ref{eq:lin_eq}).
We assume that the vector $\bm{b}$ is a real and normalized.

In the family of CG method, here the shifted COCG method,
it is important that the approximate solution of
Eq.~(\ref{Eq:ref}) is searched
within the Krylov subspace ${\cal K}_n (A,\bm{b})$.
The subspace ${\cal K}_n (A,\bm{b})$
becomes whole space at $n=N-1$,
and the solution becomes exact.

The accuracy of the approximate
solution at $n$-th iteration $\bm{x_{n}}$ is evaluated
by using the residual vector,
\begin{equation}
	\bm{r}_n = \bm{b}-A\bm{x}_n \label{Eq:residual},
\end{equation}
and the iteration is stopped
as soon as the norm of the residual vector, $||\bm{r}_n||$,
satisfies the criterion for the convergence.

The residual vectors are ``orthogonalized'' with respect to
the non-standard ``inner-product'' $(\bm{u},\bm{v}) = \bm{u}^T \bm{v}$.
When $\eta=0$, all the relevant vectors are real and the
``inner-product'' and ``orthogonality'' reduce to standard ones, respectively.
Because $\bm{r}_n$'s are ``orthogonalized'', it is convenient to use
them as base vectors of ${\cal K}_n (A,\bm{b})$.
In addition to that, owning to the ``orthogonality'', we obtain the important
theorem of ``{\it collinear residual}'' (See appendix {\ref{app:math}}).

\subsection{COCG method}
The shifted COCG method starts from the COCG method~\cite{Vorst}
solving Eq.~(\ref{Eq:ref}).
We define  $\bm{x}_n$, $\bm{p}_n$ and $\bm{r}_n$
as the approximate solution at $n$-th iteration,
the searching direction to the approximate solution at the next
iteration, and the residual vector, respectively.
At a reference energy, we must solve the following equations
under the initial conditions, $\bm{x}_0=\bm{p}_{-1}=\bm{0}$, $\bm{r}_0=\bm{b}$,
$\alpha_{-1}=1$, and $\beta_{-1}=0$:
\begin{eqnarray}
	\bm{x}_n &=& \bm{x}_{n-1} + \alpha_{n-1} \bm{p}_{n-1} \label{Eq:CG:x} , \\
	\bm{r}_n &=& \bm{r}_{n-1} - \alpha_{n-1} A \bm{p}_{n-1} \label{Eq:CG:r} , \\
	\bm{p}_n &=& \bm{r}_{n}   + \beta_{n-1} \bm{p}_{n-1} \label{Eq:CG:p} , \\
	\alpha_{n-1} &=& \frac{(\bm{r}_{n-1},\bm{r}_{n-1})}{(\bm{p}_{n-1}, A \bm{p}_{n-1})} \label{Eq:CG:alpha} , \\
	\beta_{n-1}  &=& \frac{(\bm{r}_{n},\bm{r}_{n})}{(\bm{r}_{n-1},\bm{r}_{n-1})} \label{Eq:CG:beta}.
\end{eqnarray}
Here, we must notice the fact, in the procedure of iteration, 
$(\bm{v},\bm{v})=0$ can happen though $\bm{v}\ne\bm{0}$.~\cite{inner_product}
This cannot happen in the CG method ($\eta_{\rm ref}=0$,
the matrix $A$ is positive definite) and
the other part is perfectly identical to the CG method. 
A set of residual vector $\bm{r}_n$ forms the ``orthogonalized'' base.
This ``orthogonality'' is very important for us to understand
the theorem of collinear residual.
We explain it in detail in Appendix~\ref{app:math}.

We can choose an alternative set of the recurrence equations, as follows.
Eliminating $\bm{p}$'s from Eqs.~(\ref{Eq:CG:r}) and (\ref{Eq:CG:p}),
we obtain the recurrence equation of $\bm{r}_n$,
\begin{equation}
	\bm{r}_{n+1} =
\left(1+\frac{\beta_{n-1}\alpha_{n}}{\alpha_{n-1}} -\alpha_n A\right) \bm{r}_n
 - \frac{\beta_{n-1}\alpha_{n}}{\alpha_{n-1}} \bm{r}_{n-1}. \label{Eq:shift:r}
\end{equation}
Taking ``inner product'' between $\bm{r}_n$ and the Eq.~(\ref{Eq:shift:r}),
we obtain
\begin{equation}
	\alpha_n = \frac{(\bm{r}_n,\bm{r}_n)}
	{(\bm{r}_n, A \bm{r}_n)-\frac{\beta_{n-1}}{\alpha_{n-1}}(\bm{r}_n,\bm{r}_n)}.
\label{eq:recurr}
\end{equation}
Then the Eqs.~(\ref{Eq:CG:beta}), (\ref{eq:recurr}) and (\ref{Eq:shift:r})
can produce all the base vectors, $\bm{r}_k$'s ($k > n$),
when $\alpha_{n-1}$, $\bm{r}_{n-1}$ and $\bm{r}_{n}$ are supplied.

\subsection{Shifted equations}
\label{sub:shift}
The key to the reduction of the matrix-vector operations
in solving the shifted system Eq.~(\ref{Eq:shifted}),
is the theorem of collinear residual:
\begin{equation}
	\bm{r}^\sigma_{n}=\frac{1}{\pi^{\sigma}_{n}}\bm{r}_{n}, \label{Eq:collinear}
\end{equation}
where the $\pi^\sigma$ is a scalar function (actually polynomial) of $\sigma$.
Then, once $\{\bm{r}_n\}$ are given,
the base set $\{\bm{r}_n^\sigma\}_n$ for the arbitrarily shifted system
can be obtained by using scalar multiplication.
We obtain the recurrence equations that determines
$\pi_n^\sigma, \alpha_n^\sigma,
\beta_n^\sigma, \bm{x}_n^\sigma$, and $\bm{p}_n^\sigma$,
from Eqs.~(\ref{Eq:CG:x})$\sim$(\ref{Eq:CG:beta}),
with replacing $A$ by $A+\sigma$, with the same
initial conditions:
\begin{eqnarray}
	\pi^\sigma_{n+1} &=&
 \left(1+\frac{\beta_{n-1}\alpha_{n}}{\alpha_{n-1}} + \alpha_n \sigma\right) \pi^\sigma_n
 - \frac{\beta_{n-1}\alpha_{n}}{\alpha_{n-1}} \pi^\sigma_{n-1}, \nonumber \\
	&& \label{Eq:shift:pi} \\
	\alpha^\sigma_n &=&\frac{\pi^\sigma_n}{\pi^\sigma_{n+1}}\alpha_n \label{Eq:shift:alpha},\\
	\beta^\sigma_{n}&=&\left(\frac{\pi^\sigma_n}{\pi^\sigma_{n+1}}\right)^2 \beta_n ,
    \label{Eq:shift:beta} \\
	\bm{x}^\sigma_{n} &=& \bm{x}^\sigma_{n-1} + \alpha^\sigma_{n-1} \bm{p}^\sigma_{n-1} \label{Eq:shift:x} , \\
	\bm{p}^\sigma_{n} &=& \frac{1}{\pi^\sigma_n}\bm{r}_{n} + \beta^\sigma_{n-1} \bm{p}^\sigma_{n-1} \label{Eq:shift:p}.
\end{eqnarray}
These recurrence equations can be {\bf solved without time consuming matrix-vector operation}.
In addition to that, each component of the vector
Eqs.~(\ref{Eq:shift:pi})$\sim$(\ref{Eq:shift:p})
can be solved separately,
due to the absence of the matrix operation.

\subsection{Crucial remarks for extremely large matrix
to save required memory space}

\label{sub:reconstruction}
For the solution of the relatively small matrix (${\rm Dim} \lesssim 10^{4}$),
any loop structure of shifted COCG method can be applicable.
However, in the many-electron theory,
the dimension of the intermediate vectors is huge
and then the number of intermediate vectors is restricted
to some tens or hundreds. 
In the standard loop structure which Frommer showed,~\cite{Frommer}
the outermost loop index is the iteration step $n$, and,
all the vectors $\bm{p}_n^\sigma$, $\bm{r}_n^\sigma$ and $\bm{x}_n^\sigma$
for every energy mesh point $\sigma$, are required
in order to start the calculation at the iteration step $n+1$.
Then all the energy mesh points must be fixed before the calculation starts.
Because the calculation at the respective energy mesh points are independent
to each other, the loop structure can be transformed such that
the reference system is solved with the COCG method storing the
$\{\alpha_n\}_n$,$\{\beta_n\}_n$ and $\{\bm{r}_n\}_n$,
then the shifted systems are solved with stored information about reference system
for each energy mesh points.
Here the innermost loop index is the iteration step $n$.
Since the number of energy mesh points is larger than the number of iteration generally,
the latter transformed loop structure requires smaller memory than the original one.
In addition to that, we need not to prepare 
energy mesh points $\{\sigma\}$ because all the required information
related to the reference system are stored in the COCG process,
the preceded part of the algorithm.
Then, for example, we can change the smearing factor $\eta$ freely without repeating
COCG process that includes matrix-vector operations.

Further reduction of the required memory is possible,
with further modification of the recurrence equations for the shifted system.
Assuming that a real constant vector $\bm{c}$ is an adjoint vector
and taking the inner product between $\bm{c}$ and
the Eqs.~(\ref{Eq:shift:x}) and (\ref{Eq:shift:p}), we obtain
a set of self-contained equations for determining the $(\bm{c},\bm{x}_n)$,
$n$-th approximate solution of the element of Green's function,
due to the absence of matrix-vector operations.

In the applications in Sec.~\ref{Sec:Application},
we are interested in the case where
$\bm{c}=\bm{b}$ in order to calculate the trace of the Green's function.
Therefore, we need to store $(\bm{b}, {\bm{r}_n})$, $\alpha_n$, $\beta_n$ and $||\bm{r}_n||$,
in the COCG part, and later, solve only the $\bm{b}$-component of
Eqs.~(\ref{Eq:shift:x}) and (\ref{Eq:shift:p}).
Here, the norm of the residual vector
$||\bm{r}^\sigma_n||=\frac{1}{|\pi^\sigma_n|}||\bm{r}_n||$
is not necessary to solve the recurrence equations but
is used to monitor the convergence of the approximate solution.
Additionally, we store the full components of
the last two $\bm{r}_n$'s in the COCG part,
in order to extends the iteration number in the seed switching part
(subsection~\ref{sub:seed_switch}).

Even if the full components of the Green's function are needed,
we need to store just a few components of ${\bm{r}_n}$,
because the suffix of the Green's function denotes the one-electron orbitals,
the number of which is very small compared to the dimension of many-electron Hamiltonian $H$.

\subsection{Preparation of ground state wavefunction}
\label{sub:eigen}

The transformation of the loop structure
in the subsection~\ref{sub:reconstruction} increase
the re-usability of the program code.
The COCG part of the code can also be used in
the process to improve the ground state wavefunction as follows.

First we use the Lanczos method in order to tridiagonalize the Krylov subspace,
then, obtain the ground state diagonalizing it.
The calculated lowest eigenenergy converges rapidly
with increase of the dimension of the subspace,
but the wavefunction does not,
due to the unstable orthogonality against the inevitable rounding error.
Next we improve the approximate eigenenergy and the wavefunction
with the inverse iteration method.
Because the COCG process with the real arithmetics
is the same as CG process,
here we can use the COCG part of the shifted COCG algorithm
whose loop structure is changed as
in the subsection~\ref{sub:reconstruction}.~\cite{COCG_real}

Since the inverse iteration method works only when the approximate
eigenvalue and eigenvector are given,
the first Lanczos process can not be omitted.
If the accuracy of the calculated wavefunction is not enough,
the processes are repeated with replacing the initial Lanczos
vector by the latest approximate wavefunction.

\section{Accuracy and seed switching}

\label{Sec:error_and_seed_switching}

\subsection{Estimating accuracy of Green's function}
\label{sec:numerical_error}

In this subsection, we explain the accuracy of the Green's function calculated
by the shifted COCG method and give its estimation.
Here $G_{\rm exact}$ is the ``exact'' solution
of Eqs.~(\ref{eq:lin_eq}) and (\ref{eq:gf2}) for a given finite value of $\eta$.
Then we say that ``the calculated Green's function is accurate'',
when the $|\frac{[G_{\rm sCOCG}-G_{\rm exact}]_{ij}}{[G_{\rm exact}]_{ij}}|$
(hereafter, ``accuracy'') is small.
The ``accuracy'' and $||\bm{r}_n||$ are generally ``truncation error''
of $G$ and $\bm{x}_n$, respectively.
Because the shifted system is equivalent to the
reference system, then we can estimate the accuracy for
the shifted system,
with replacing $A$ by $A+\sigma$ and any other quantities $\{q\}$
by $\{q^\sigma\}$.

We can derive following equation from
Eqs.~(\ref{eq:lin_eq}),(\ref{eq:gf2}) and (\ref{Eq:residual}),
\begin{equation}
   \left|\frac{[G_{\rm sCOCG}-G_{\rm exact}]_{jj}}{[G_{\rm exact}]_{jj}}\right|
 = \left|\frac{(\bm{b}, A^{-1} \bm{r}_n)}{(\bm{b}, A^{-1} \bm{b})}\right| .
	 \label{Eq:Truncation}
\end{equation}
If the matrix $A$ was positive definite real symmetric matrix and
the vector $\bm{b}$ and $\bm{r}_n$ were real vectors,
the upper bound of the right hand side of Eq.~(\ref{Eq:Truncation})
is equal to $\frac{||\bm{r}_n||}{||\bm{b}||}=||\bm{r}_n||$.

When the matrix $A$ can be fully diagonalized numerically,
we can estimate $G_{\rm exact}$
within rounding errors, and then,
obtain the ``accuracy'' of the approximate Green's function calculated
by shifted COCG method.

\begin{figure}
    \begin{center}
    \caption{
	An example of the ``accuracy'' of $[G_{\rm sCOCG}]_{jj}$,
	$||\bm{r}_n||$  and the imaginary part of the ``exact''
	solution ${\rm Im} [G_{\rm exact}]_{jj}$ (See text).
	The dimension of the matrix $A$ here is equal to 8,960,
	in the same model in Sec.~\ref{Sec:Application}.
	The reference energy and smearing factor are equal to $\omega_{\rm ref}=6.04$eV
	and $\eta=0.05eV$. The iteration number is equal to 800.
	}
    \label{Fig:error}
  \end{center}
\end{figure}

The dotted and solid line in the Fig.~\ref{Fig:error} show
the ``accuracy'' of $[G_{\rm sCOCG}]_{jj}$ and the norm of the
residual vector $||\bm{r}_n||$, respectively.
The dashed line shows the excitation spectra.
The figure shows that
the ``accuracy'' is bounded by the norm of the residual vector.
Therefore, we can estimate the ``accuracy'' of the calculated
Green's function by using $||\bm{r}_n||$, without the knowledge of
the ``exact'' solution $G_{\rm exact}$.
We can also see from the figure that 
the Green's function calculated by the shifted COCG method
is accurate more, near the bounds of the spectra.

\subsection{Seed switching}
\label{sub:seed_switch}

Assuming that the approximate solution of the reference system,
Eqs.~(\ref{Eq:CG:x})$\sim$(\ref{Eq:CG:beta}),
converges at $M$-th iteration,
we can solve the shifted system,
Eqs.~(\ref{Eq:shift:pi})$\sim$(\ref{Eq:shift:p}),
up to the same $M$-th iteration.
When the approximate solution of the shifted system
does not converges at any $\omega+{\rm i}\eta=\omega_{\rm ref}+{\rm i}\eta_{\rm ref}+\sigma$,
we should extend the iteration of the reference system.
However at $\omega_{\rm ref}+{\rm i}\eta_{\rm ref}$,
the extension does not improve the approximate solution,
since the norm of the residual vector is considerably small already.
In that occasion,
we should change the seed ($\omega_{\rm ref}+{\rm i}\eta_{\rm ref}$) to a new one,
$\omega_{\rm ref}^{\rm new}+{\rm i}\eta_{\rm ref}^{\rm new}$,
where the norm of the residual vector is large and the approximate
solution does not converge.
Because the shifted system is equivalent to the reference system,
we can change the seed as follows,
without disposing the previous calculation at the old $\omega_{\rm ref}$.~\cite{Sogabe}


We define $\sigma_{\rm max}$ so that $\bm{r}_M^{\sigma_{\rm max}}={\rm Max}_{\sigma} \{\bm{r}_M^\sigma\}$,
where ${\rm Max}_{\sigma}$ means the maximum value on the $\sigma$-mesh
(energy mesh) points.
Because the $\omega_{\rm ref}+{\rm i}\eta_{\rm ref}+\sigma_{\rm max}$ is
the prime candidate for the energy of the slowest convergence,
we choose it as the $\omega_{\rm ref}^{\rm new}+{\rm i}\eta_{\rm ref}^{\rm new}$.
Then, $\alpha_{k}^{\sigma_{\rm max}}, \beta_{k}^{\sigma_{\rm max}}$, and
$\bm{r}_{k}^{\sigma_{\rm max}}$ ($k=0,1,\cdots,M$) are calculated and
replaces the old values at old reference energy, respectively.
Finally, the recurrence Eqs.~(\ref{Eq:CG:x})$\sim$(\ref{Eq:CG:beta})
of the COCG method at $\omega_{\rm ref}^{\rm new}+{\rm i}\eta_{\rm ref}^{\rm new}$
are calculated until the solution converges at $M^{\rm new}$-th iteration.
\begin{figure}
    \begin{center}
    \caption{The seed
	($\omega_{\rm ref}+{\rm i}\eta_{\rm ref}$)
	switching and the norm of residual vector
	$||\bm{r}_n||$ at respective seed (the black solid line).
	Here the $\eta_{\rm ref}$ is a constant and equal to $0.05$eV.
	The dashed vertical line indicates the iteration number where the seed is switched.
	The gray lines show the $||\bm{r}_n^{\sigma_{\rm max}}||$'s (see text).
	The calculated values of $||\bm{r}_n||$ at $\omega_{\rm ref}=8.95$eV
	are multiplied by the factor of $100$ in order to avoid overlap.
	}
    \label{Fig:seed_switch}
  \end{center}
\end{figure}
The important point of the seed switching is that we can recalculate new
$\alpha_n$'s, $\beta_n$'s, and $||\bm{r}_n||$'s ($0 < n \leq M$)
without any matrix-vector operation,
though the matrix-vector operation is required to calculate
the new ones at further $n'$-th iteration ($M < n' \leq M^{\rm new}$).

The same remark as the subsection \ref{sub:reconstruction} is applicable
to the implementation of the seed switching.
The shifted COCG algorithm with seed switching of any loop structure
can be applicable to relatively small matrices, but,
we must change the loop structure of it
from the previous one~\cite{Sogabe},
since the size of the intermediate vector is huge in the many-electron
theory.
We must even change the recurrence equation of $\bm{r}_n$ to
Eq.~(\ref{Eq:shift:r}),
so that the $\bm{r}_{n+1}$ is calculated only with
the $\bm{r}_{n-1}$ and $\bm{r}_n$,
in stead of all $\{\bm{r}_n\}_{n=1,2,\ldots,n}$,
for extremely huge matrices.
When we store $\{\alpha_n,\beta_n\}_{n=0,1,\ldots,M-1}$,
$\bm{r}_{M-1}$ and $\bm{r}_{M}$ in the COCG process,
then we can calculate $\alpha_{M-1}^{\sigma_{\rm max}}$,
$\bm{r}_{M-1}^{\sigma_{\rm max}}$ and $\bm{r}_{M}^{\sigma_{\rm max}}$,
which are required for the following calculations.
Then we calculate the new reference system
up to $M^{\rm new}$-th iteration step,
by using Eqs.~(\ref{Eq:CG:beta}), (\ref{eq:recurr}) and (\ref{Eq:shift:r}).

Figure~\ref{Fig:seed_switch} shows the example of the seed switching.
The system is the same one as in Fig.~\ref{Fig:error} except $\omega_{\rm ref}$.
Here the $\eta=\eta_{\rm ref}=0.05$eV.
At the $\omega_{\rm ref}=-1.40$eV, $||\bm{r}_n||$ decrease exponentially,
and satisfies the criterion $||\bm{r}_n||< 10^{-10}$ at iteration step $n=330$.
However, there are many energies where the converging speed of the residual
vector $||\bm{r}^\sigma_n||$ is slower than that at $\omega_{\rm ref}$.
Then the $\sigma_{\rm max}=7.43$eV is searched and the
$\omega_{\rm ref}^{\rm new}$ is shifted to be $6.03$eV.
We need seed switching twice more at $1509$- and $3297$-th iteration
in order to obtain the global convergence.
The largest value of the ``accuracy'' is $1.5 \times 10^{-10}$
at the last iteration step $n=3455$.

\subsection{Robustness of shifted COCG method}

In this subsection, we explain the
robustness of the shifted COCG method,
which is very important to obtain the converged approximate
solution, especially in the case of the long iteration.
We say that the calculation is robust,
when the calculation is stable against the perturbation.
For examples, the orthogonality of $\{\bm{r}_n\}$ is
not a robust property, because
the inevitable rounding error perturb
the calculation and the orthogonality is broken down quickly.

The robustness of the shifted COCG method consists of
two parts. One is the robustness of COCG method
at the reference energy $\omega_{\rm ref}$.
And the other is the robustness of the iterative solution of
the shifted equations.
Figure~\ref{Fig:seed_switch} shows the robustness of the COCG method
at $\omega_{\rm ref}$, because the norm of the residual vector
$||\bm{r}_n||$ goes to 0 in spite of long iteration 3,540.
The global convergence of the ``accuracy'' that is mentioned at the end of
the subsection~\ref{sub:seed_switch} shows 
the robustness of the iterative solution of the shifted equations.
In the shifted COCG method, the ``orthogonality'' of base vectors $\{\bm{r}_n\}$
is not necessary for reducing  $ ||\bm{r}_n||$,
in contrast to the fact that
the subspace diagonalization methods requires the unitarity of
the base vectors.




\section{Application of the shifted COCG method to the many-electron problem}
\label{Sec:Application}

Here we apply the shifted COCG method, to the
double orbital extended Hubbard Hamiltonian
and calculate the excitation spectra.~\cite{Yamamoto}

\subsection{Hamiltonian of La$_\frac{3}{2}$Sr$_\frac{1}{2}$NiO$_{4}$}

The experimental results show that the
layered perovskite La$_\frac{3}{2}$Sr$_\frac{1}{2}$NiO$_{4}$ 
is an insulator with charge and spin stripe order,
as depicted in Fig.~\ref{Fig:ChargeSpinOrder}.
The charge and spin structures of the single layer of La$_\frac{3}{2}$Sr$_\frac{1}{2}$NiO$_{4}$
(pseudo two-dimensional system), choosing Ni 3d $e_g$ orbitals as relevant ones, 
was studied with the extended Hubbard model recently.~\cite{Yamamoto}
Here we use the same Hamiltonian.
\begin{eqnarray}
\hat{H} &=& \sum_{\scriptstyle i,j, \alpha,\beta, \sigma}
  t_{i \alpha j \beta}   \hat{c}^{\dagger}_{i \alpha \sigma}
                         \hat{c}_{j \beta \sigma}
 + \sum_{i, \alpha, \sigma}
  \varepsilon_{i \alpha} \hat{n}_{i \alpha \sigma} \nonumber \\
&+& U      \sum_{i,\alpha} \hat{n}_{i \alpha \uparrow}\hat{n}_{i \alpha \downarrow}
+ (U-2J) \sum_{i,\sigma,\sigma'}\hat{n}_{i ,3z^2-1, \sigma}\hat{n}_{i ,x^2-y^2, \sigma'} \nonumber \\
&+& \frac{J}{2} \sum_{\STACK{i,\alpha\ne\beta,}{\sigma,\sigma'}}
\left(
\hat{c}^{\dagger}_{i \alpha \sigma} \hat{c}^{\dagger}_{i \beta \sigma'} 
          \hat{c}_{i \alpha \sigma'}          \hat{c}_{i \beta \sigma}
+
\hat{c}^{\dagger}_{i \alpha \sigma} \hat{c}^{\dagger}_{i \alpha \sigma'} 
          \hat{c}_{i \beta  \sigma'}          \hat{c}_{i \beta \sigma} \right)
 \nonumber \\
&+& V \sum_{\STACK{\left<i,j\right>,\alpha,}{\beta, \sigma, \sigma'}}
\hat{n}_{i \alpha \sigma} \hat{n}_{j \beta \sigma'} ,  \label{eq_reduced_hamiltonian}
\end{eqnarray}
where the suffix $\{i,j\}$ denote the site, 
$\{\alpha,\beta\}$ denote the orbital $3z^2-1$ or $x^2-y^2$,
and $\{\sigma,\sigma'\}$  denote the spin co-ordinate.
The annihilation and number operator are $\hat{c}$ and $\hat{n}$, respectively.
The symbol $t$, $\varepsilon$, $U$, $J$, $V$ denote
the Slater-Koster type hopping parameter, single electron energy, on-site Coulomb
interaction, on-site exchange interaction, intersite Coulomb interaction, respectively.
Hopping parameters are finite for nearest neighbor (n.n.) and second n.n. pair of sites.
The braces $\langle\cdots\rangle$ means that two sites enclosed by them
are the n.n. sites.

Though the anisotropy of the hopping parameters for the second n.n. pair stabilizes
the spin structure,~\cite{Yamamoto} we choose the isotropic (tetragonal) parameter set
shown in Table~\ref{Tab}.
The role of $U$ and $J$ in the present situation is stabilization
of integral valency of Ni ions (Ni$^{3+}$ and Ni$^{2+}$) and spin polarization.

\begin{table}[h]
	\begin{tabular}{cccccccc}
		$t_{dd\sigma}$	&	$t_{dd\delta}$	& $\frac{1}{4}t'_{dd\sigma} + \frac{3}{4}t'_{dd\delta}$ &
	 	$t'_{dd\pi}$	&	$\Delta$	& $U$ 	& $J$ &	$V$ \\
 \hline
		-0.543 			& 0.058 & -0.018 & -0.023 & 0.97 & 7.5 & 0.88 & 0.5
\end{tabular}
	\caption{The values of parameters in the Hamiltonian in unit of eV.~\cite{Yamamoto}
	The prime symbol at the right shoulder of ``$t$'' means the second n.n. hopping.
	}
	\label{Tab}
\end{table}

\begin{figure}[hbt]
\begin{center}
	\caption{Experimentally observed charge and spin order in the La$_\frac{3}{2}$Sr$_\frac{1}{2}$NiO$_4$.
	\cite{Yoshizawa_2000}}
	\label{Fig:ChargeSpinOrder}
\end{center}
\end{figure}



\subsection{Ground state of La$_\frac{3}{2}$Sr$_\frac{1}{2}$NiO$_{4}$}

Here we summarize the properties of the calculated ground state
of the La$_\frac{3}{2}$Sr$_\frac{1}{2}$NiO$_{4}$.~\cite{Yamamoto}
The calculated ground state shows the charge and spin stripe order 
consistent with experimental observation and the system is insulator.
Diagonal hole stripes are separately localized on Ni$^{3+}$ site in order to 
reduce hole-hole interaction energy induced by inter-site Coulomb interaction $V$. 
Charge order and the inter-site Coulomb interaction $V$ are directly related to 
the energy gap in the excitation spectra of the system.
Spin stripe occurs only under the condition of the existence of multi-orbitals
and the charge order with a help of anisotropy.
The spin stripe is determined by the electronic structure with smaller energy scale 
than that of the charge stripe. 

\subsection{Computational details}
Here we explain miscellaneous computational details.

The calculated system is two-dimensional square lattice.  
There are 12 electrons on the periodic $\sqrt{8}\times\sqrt{8}$ sites.
Because the $\mbox{total }S_z$ of the system is preserved, 
we can reduce the number of relevant many-electron states to 64,128,064,
by using the condition $S_z =0$.

The smearing factor $\eta$ is also an arbitrary parameter in the present paper.
Here we explain how we chose the value of $\eta$.
The energy scale of the low energy excitations is
$t \sim V \sim {\rm O}(10^{-1} eV)$, because the value of
on-site Coulomb interaction $U$ is much larger.
Therefore we must set $\eta$ lower than $10^{-1}$eV,
so that $\eta$ does not smear out the finer structure of the spectra than itself.
There is another restriction that the interval of the energy mesh
is small enough than $\eta$, in order to see
the fine peak structure of the spectra.
Then, because the calculation time increases with decreasing $\eta$,
the value of $\eta$ is roughly determined as O(10$^{-3}$eV)$\sim$O(10$^{-2}$eV).

Next point is the criterion for the convergence of the ground state vector.
Our calculations are of the double precision and the rounding error is inevitable
($\sim 10^{-16}$) in the each component of the eigenvector.
Assuming the accumulated error is of O($\sqrt{N}$) ($N=$64,128,064),
the accuracy is expected to be $10^{-16}\times\sqrt{N}\sim 10^{-12}$.
We set the allowance for the estimation by factor of $10^2$,
and the criterion for the accuracy of calculated ground state energy $E_{\rm gs}$ and
eigenvector $\left| \right\rangle$ is 
$\sqrt{\left\langle \right| (\hat{H}-E_{\rm gs})^2 \left| \right\rangle}< 10^{-10}$.

\begin{figure}
\begin{center}
	\caption{The spectral functions of the present
	Hamiltonian of double orbital extended Hubbard model with
	12 electrons on the periodic $\sqrt{8}\times\sqrt{8}$ sites (gray line)
	and their ``accuracy'' (black line, see text).
	The spectra of affinity and ionization levels are calculated separately
	by using the shifted COCG method, for the given smearing factor $\eta=0.01$eV
    (upper panel) and $\eta=0.10$eV (lower panel).
	The highest occupied level is at $9.4$eV and lowest unoccupied $10.3$eV.
	Intersite Coulomb interaction $V=0.5$eV.
	Energy zeroth is set at the ground state energy $36.755$eV of 12 electron system.}
	\label{Fig:eta}
\end{center}
\end{figure}

\subsection{Spectral function}

We examined the spectral function
\begin{equation}
{\cal A}(\omega)=-\frac{1}{\pi}{\rm Tr}\left[ {\rm Im} G(\omega) \right].
\end{equation}
This can be easily evaluated by the shifted COCG method.  
Figure~\ref{Fig:eta} shows the spectral functions of the state D at $V=0.5$eV.
The upper and the lower panel show the case of $\eta=0.1$eV and $\eta=0.01$eV, respectively.
Both of them are calculated from the same COCG calculations
and the only difference between them is the imaginary part of the energy shift $\sigma$.
The spectra of ionization and affinity levels are calculated separately and
the $\omega_{\rm ref}$'s for respective spectra are chosen to be $(9.0+{\rm i} 0.01)$eV
and  $(10.4+{\rm i} 0.01)$eV.
The highest occupied level is at $9.4$eV and lowest unoccupied $10.3$eV.
The number of iterations equals to 800 for each spectra.
If one attempt to obtain the profile of the spectra with smoothly connected curves
as is in the bulk limit,
one should set the value of $\eta$ sufficiently larger than
$\frac{(\mbox{width of spectra})}{(\mbox{iteration number})}$,
in order to smear out the excessive peaks caused by the finite system.
Because iteration number equals to 800 in the present calculations,
this criterion becomes $\eta \gg 0.03$eV,
and the gray curves in the upper panel of Fig.~\ref{Fig:eta} shows
the smooth profile of the spectral function.
If one attempt to see whether the energy gap opens at the boundary between
affinity and ionization levels, one must choose sufficiently smaller $\eta$ than
the width of energy gap.
In the present calculation, this criterion becomes  $\eta \ll 0.9$eV, 
and, the gray curves in the lower panel of Fig.~\ref{Fig:eta} show
the energy gap around $\omega=9.8$eV.
We can choose $\eta$ independent with the reference energy,
then the energetic resolution of the spectral function can be changed
after all the time consuming matrix-vector operations have been finished.


The black curves in the upper and lower panels of the Fig.~\ref{Fig:eta}
show the ``accuracy'' of the respective spectral functions, and
the spectra are extremely accurate near the boundary of the spectra
($\omega=9.8$eV), where the energy gap is open.
Therefore, we conclude from the lower panel of Fig.~\ref{Fig:eta}
that the ground state of the present Hamiltonian is insulator.

Changing the value of $V$ continuously from $0.5$eV to $0.0$eV,
we find that the system becomes metal.~\cite{Yamamoto}
Therefore, the intersite Coulomb interaction makes the present system insulator,
unlike the usual transition metal oxide
where the large on-site Coulomb interaction makes the system insulator.

\section{Discussion and summary}
\label{Sec:Discussion}
Once COCG method is applied to the reference system 
Eq.~(\ref{Eq:ref}), the shifted system Eq.~(\ref{Eq:shifted}) is
solved without time consuming matrix-vector operations,
by shifted COCG method.
This notable property is due to the mathematical structure of COCG method,
such that the residual vector's are forming the ``orthogonal''
base set of vectors, whose direction does not change against $\sigma$.
This reduction of the matrix-vector operation extremely
accelerate the calculation speed of Green's function
$G(\omega)$, keeping the robustness of COCG method.
Simultaneously, the accuracy of the approximate $G(\omega)$ is easily
estimated as the norm of the newly generated base vector (residual vector)
at the latest iteration.
The total accuracy of the shifted COCG method varies
depending on $\omega$, $\sigma$ and
$\omega_{\rm ref}$~\cite{note_for_omega_ref} and
generally very small near the bounds of the respective spectra.
In the many-electron Green's function,
we are usually interested in the low energy excitations,
in other words,
the spectra near the boundary between affinity and ionization levels.
Therefore, we can calculate the Green's function
accurately and quickly by the shifted COCG method, in the interesting energy range.

Another problem in the application of the COCG method to the
many-electron theory is a memory constraint due to the extremely large size
of vectors and matrices.
We resolved this problem with separating 
the COCG part for the reference system
and the part for the shifted equations, changing the loop structure.
This change give us the following two merits.
One is a usage of the former part for improving the ground state,
as is mentioned in the subsection~\ref{sub:eigen}.
The other is the fast calculation of changing smearing factor $\eta$,
which is just a imaginary shift.
When we do not know the proper energy scale {\it a priori},
the width of the energy gap in the present paper
or the proper value of $\eta$, this merit is very important.

The seed switching is a very important idea
for the shifted COCG method to give global convergence,
which means that the calculated solution converges
everywhere in the interested energy region.
Because it takes much iteration steps to converge the solution
especially in the middle of the spectra,
sometimes we must discontinue the iteration step before obtaining global convergence.
In that case, we must examine the accuracy of the result and check if
the solutions in the required energy range satisfy the criterion.

The applicability of the above reconstruction
and the seed switching are not specific to 
the many-electron problem.
We can apply them to the general solution of
the Green's function of extremely large dimension.

We applied the shifted COCG method to the charge and spin order
in La$_\frac{3}{2}$Sr$_\frac{1}{2}$NiO$_{4}$,
where the intersite Coulomb interaction, relatively small compared with on-site one,
plays an important role.
Then we conclude the relatively small energy gap opens at the Fermi energy,
and the system becomes insulator, due to the intersite Coulomb interaction.

\begin{acknowledgments}

Calculations were done at the Supercomputer Center, Institute for Solid State 
Physics, The University of Tokyo.
This work was partially supported by a Grant-in-Aid for Scientific Research in Priority
Areas ``Development of New Quantum Simulators and Quantum Design'' (No.170640004) of
The Ministry of Education, Culture, Sports, Science, and Technology, Japan.

\end{acknowledgments}

\appendix

\section{Mathematical structure of shifted COCG methods}
\label{app:math}
Here we explain the two important points to understand
the mathematical structure of the shifted COCG methods.~\cite{Zhang,Frommer}

One is the ``orthogonality'' of the residual vectors $\{\bm{r}_k\}$
with respect to a non-standard ``inner-product'' $(\bm{u},\bm{v})=\bm{u}^T \bm{v}$,
which the theorem of collinear residual is based on.
On should be noticed that this ``orthogonality''
is different from the well-known ``$A$-orthogonality''
of the searching directions $\{\bm{p}_k\}$.
Because $\{\bm{r}_k\}_{k=0,1,\ldots,n-1}$ is a base set of ${\cal K}_{n-1}(A,\bm{b})$,
the ``orthogonality'' is also represented as follows
\begin{equation}
	\bm{r}_n \in   {\cal K}_{n}(A,\bm{b}) \mbox{ and }
    \bm{r}_n \perp \overline{{\cal K}_{n-1}(A,\bm{b})}, \label{Eq:Krylov_perp}
\end{equation}
where the over line stands for taking conjugate.
Therefore, the direction of $\bm{r}_{n}$ is {\bf uniquely} determined
by above equation, for arbitrary $A$, as the 1-dimensional complementary
space of ${\cal K}_{n-1}(A,\bm{b})$ within ${\cal K}_{n}(A,\bm{b})$.
Then the theorem of the collinear residual Eq.~(\ref{Eq:collinear})
is derived from the invariance of Krylov subspace against $\sigma$.

Another point appears in the similarity between the two
Eqs.~(\ref{Eq:shift:r}) and (\ref{Eq:shift:pi}).
Representing ${\bm{x}_n}$ and ${\bm{r}_n}$ as
$X_{n-1}(A)\bm{b}\in {\cal K}_{n-1}(A,\bm{b})$ and
$R_n(A)\bm{b}\in {\cal K}_{n}(A,\bm{b})$, respectively, 
we can derive the relation between two polynomials $R_n(t)=1-t X_{n-1}(t)$
from Eq.~(\ref{Eq:residual}).
This relation and the theorem of the collinear residual lead to the relation
between the two polynomials $\pi^\sigma_{n} = R_{n}(-\sigma)$.~\cite{Frommer}
This relation explains the similarity between the two
Eqs.~(\ref{Eq:shift:r}) and (\ref{Eq:shift:pi}),
and the plus sign in front of $\sigma$ in the latter equation.

Thus the mathematical structure of the shifted COCG method consists of
two structures, that of a vector space and that of a set of polynomials.
The non-standard ``inner product'' in the shifted COCG method can be recognized
as the conservation of the analytic property as a polynomial of $\sigma$.

\section{Reducing storage size of Hamiltonian}
\label{app:storage}
In investigating the properties of the many-electron Hamiltonian,
there is a trade-off between the speed of matrix-vector operation
and the amount of the memory where the matrix elements of the Hamiltonian
are stored.
Assuming the all non-zero matrix elements are stored separately
and the number of the non-zero matrix elements per each column is equal to 20,
then about 10 GB is required to store the Hamiltonian in the case of present paper.
That is too much for the most of modern computers.
In stead, if the operation of the Hamiltonian on the vectors is implemented as
the summation of the respective term in Eq.~(\ref{eq_reduced_hamiltonian}),
then the number of operator equals to 640, and,
the memory required to store 640 operators is
negligibly small compared to 10GB.
However, these 640 operators must be applied to the vectors,
for single operation of the Hamiltonian.
Therefore, the calculation time increases,
compared to the case of the all elements of the Hamiltonian are stored.

For simplicity, the Fermion sign and the two particle operators in the Hamiltonian
are neglected hereafter,
then, the difference between above two manners of storing the Hamiltonian
are described as follows, mathematically.
We define $V$ as a whole vector space of the single electron state
and decompose it into the direct sum $V=\oplus_{\alpha}V_{\alpha}$,
$\alpha$ denotes orbital or spin or arbitrary combinations of
relevant quantum number.
Consequently, the whole space where $n$-body Hamiltonian
acts is described as
\begin{equation}
\underbrace{V \otimes \ldots \otimes V}_{n}
= \oplus_{\alpha_1,\ldots,\alpha_n}
 \underbrace{V_{\alpha_1}\otimes\ldots\otimes V_{\alpha_n}}_{n}.
\end{equation}
Decomposing single particle operator $P$ 
as $P=\sum_{\alpha,\beta}P_{\alpha\beta}$,
$P_{\alpha\beta} V_{\beta} \in V_{\alpha}$,
then the action of $P$ on the $\otimes^{n}V$ is decomposed as
\begin{eqnarray}
P&=& (\sum_{\alpha,\beta}P_{\alpha\beta}\otimes 1 \otimes \ldots \otimes 1) + \ldots \nonumber \\
 &+& (1\otimes \ldots \otimes 1 \otimes \sum_{\alpha,\beta}P_{\alpha\beta} \otimes 1 \otimes \ldots \otimes 1) + \ldots \nonumber \\
 &+& (1\otimes \ldots \otimes 1 \otimes \sum_{\alpha,\beta}P_{\alpha\beta}), \label{Eq:partition_of_P}
\end{eqnarray}
since the single particle operator $P$ acts on $\otimes^{n}V$
as $(P\otimes 1 \otimes \ldots \otimes 1) + \ldots + (1\otimes \ldots \otimes 1
\otimes P \otimes 1 \otimes \ldots \otimes 1)+ \ldots + (1\otimes \ldots \otimes 1
\otimes P) $.
This decomposition is trivial but one should be noticed
that the dimension of respective subspace decrease exponentially with $n$ as
$\dim(V_{\alpha_1}\otimes\ldots\otimes V_{\alpha_n})=(\frac{\dim{V}}{m})^n$,
where $m$ is the number of the partition $V=\oplus_{\alpha}V_{\alpha}$.
In contrast, the number of the terms to be summed up increases in proportion to $nm^2$
Therefore, the total number of the matrix elements decrease by this decomposition in
Eq.~(\ref{Eq:partition_of_P}).
Extreme limit of the respective $V_{\alpha}$ consists of only one dimension
is corresponding to the above mentioned case of 640 operators.

In the present example of the spectral function,
we choose the decomposition $V=V_{\uparrow}\oplus V_{\downarrow}$,
because total $S_z$ is preserved, and consequently,
the numbers of $\uparrow$- and $\downarrow$-electrons are preserved,
and more, the hopping part of the present Hamiltonian
does not have the cross term with respect to spin.
This partition of the vector space leads to more simplified 
partition of the operator than Eq.~(\ref{Eq:partition_of_P}),
$H=H_{\uparrow\uparrow} \otimes 1 + 1 \otimes H_{\downarrow\downarrow} 
+(\mbox{residual part})$,~\cite{Yamada} where
$H_{\uparrow\uparrow}$ and $H_{\downarrow\downarrow}$ are including
the hopping term with respect to each spin and
a part of on-site Coulomb interaction.
As a result, the size of the memory area where the values of the matrix elements
are stored is about 1GB. Actually, an extra 0.5GB is required for storing the
indexes of the place of non-zero elements, then, totally 1.5GB is required
for storing the whole Hamiltonian.
That is not so big for a modern computer.



\begin{thebibliography}{99}

\bibitem{Tokutra}
T. Kimura, T. Goto,
H. Shintani, K. Ishizaka,
T. Arima and Y. Tokura, Nature {\bf 426}, 55 (2003).
N. Kida, Y. Kaneko, J. P. He, M. Matsubara,
H. Sato, T. Arima, H. Akoh, and Y. Tokura,
Phys. Rev. Lett. {\bf 96}, 167202 (2006).

\bibitem{Frommer}
W. A. Frommer, Computing {\bf 70}, 87 (2003).

\bibitem{Takayama}
R. Takayama, T. Hoshi, T. Sogabe, S.-L. Zhang, and T. Fujiwara,
Phys. Rev. {\bf B73}, 165108 (2006).

\bibitem{Vorst}
H. A. van der Vorst, J. B. M. Melissen, IEEE Trans. Mag. {\bf 26}, 706 (1990).


\bibitem{inner_product}
The inequivalence of $(\bm{v},\bm{v})=0$ and $\bm{v}=\bm{0}$ causes
that the $\alpha_{n}$ or $\beta_{n-1}$ may be equal to
$0$ without satisfying $\bm{r}_{n}=\bm{0}$.
When this occurs before the approximate solution 
converges, we fails to obtain the approximate solution.
However, we seldom experience such a situation.


\bibitem{Sogabe}
T. Sogabe, T. Hoshi, S.-L. Zhang, and T. Fujiwara,
``{\it Frontiers of Computational Science:
Proceedings of the International Symposium
on Frontiers of Computational Science 2005}''
edited by Y. Kanada, H. Kawamura, and M. Sasai
(Springer-Verlag, Berlin, 2007).
arXiv:math/0602652v1.
%

\bibitem{COCG_real}
When the all the matrix and vectors in the COCG part is real,
in the $\eta=0$ case,
we had better use the real matrix-vector operations,
because the calculation time is less than half of the complex
calculation time.

\bibitem{Yamamoto}
S. Yamamoto, T. Fujiwara, and Y. Hatsugai,
Phys. Rev. {\bf B76}, 165114 (2007).

\bibitem{Yoshizawa_2000} H. Yoshizawa, T. Kakeshita, R. Kajimoto, T. Tanabe, T. Katsufuji, and Y. Tokura, Phys. Rev. {\bf B61}, R854 (2000).

\bibitem{note_for_omega_ref} In the ideal calculation without numerical error,
its results do not depend on
the value of the reference energy $\omega_{\rm ref}$.
However, in the actual calculation, they depend on the $\omega_{\rm ref}$,
especially when the smearing factor $\eta$ is small.


\bibitem{Zhang} S. Fujino and S.-L. Zhang, {\it ``Hanpukuhou no suuri''}
(Written in Japanese. The title means ``{\it Elements of Iterative Methods}'')
(Asakura, Tokyo, 1996).

\bibitem{Yamada}
S. Yamada, T. Imamura, M. Machida,
ACM/IEEE SC 2005 Conference (SC'05), 44 (2005).

\end{thebibliography}
\end{document}